# Frustrated magnetic interactions in FeSe

Yiqing Gu[1,2], Qisi Wang[1,3], Hongliang Wo[1,2], Zheng He[1], Helen C. Walker[4], Jitae T. Park[5], Mechthild Enderle[6], Andrew D. Christianson[7], Wenbin Wang[8], and Jun Zhao[1,8,2,9,*]

[1]*State Key Laboratory of Surface Physics and Department of Physics, Fudan University, Shanghai 200433, China*

[2]*Shanghai Qi Zhi Institute, Shanghai 200232, China*

[3]*Physik-Institut, Universität Zürich, Winterthurerstrasse 190, CH-8057 Zürich, Switzerland*

[4]*ISIS Facility, Rutherford Appleton Laboratory, STFC, Chilton, Didcot, Oxon OX11 0QX, United Kingdom*

[5]*Heinz Maier-Leibnitz Zentrum (MLZ), Technische Universität München, 85748, Garching, Germany*

[6]*Institut Laue-Langevin, Avenue des Martyrs, CS 20156, F-38042 Grenoble Cédex 9, France*

[7]*Materials Science & Technology Division, Oak Ridge National Laboratory, Oak Ridge, Tennessee 37831, USA*

[8]*Institute of Nanoelectronics and Quantum Computing, Fudan University, Shanghai 200433, China*

[9]*Shanghai Research Center for Quantum Sciences, Shanghai 201315, China*

Abstract

The structurally simplest high-temperature superconductor FeSe exhibits an intriguing superconducting nematic paramagnetic phase with unusual spin excitation spectra that are different from typical spin waves; thus, determining its effective magnetic exchange interactions is challenging. Here we report neutron scattering measurements of spin fluctuations of FeSe in the tetragonal paramagnetic phase. We show that the equal-time magnetic structure factor, $\mathcal{S}(\mathbf{Q})$, can be effectively modeled using the self-consistent Gaussian approximation calculation with highly frustrated nearest-neighbor ($J_1$) and next-nearest-neighbor ($J_2$) exchange couplings, and very weak further neighbor exchange interaction. Our results elucidate the frustrated magnetism in FeSe, which provides a natural explanation for the highly tunable superconductivity and nematicity in FeSe and related materials.

High-temperature superconductivity emerges when the antiferromagnetic order of the parent compound is suppressed through chemical doping or applying pressure [1,2]. Strong spin fluctuations, which are closely coupled with superconductivity, persist in superconductors without any magnetic order [1-3]. Determining the effective magnetic exchange interactions of high-temperature superconductors is not only crucial to elucidate whether or how magnetism could promote superconductivity but also to understand their exotic properties in the normal states. Magnetic exchange interactions can be estimated by fitting the spin excitation spectrum using linear spin wave theory assuming that superconductors retain magnetic structures similar to their parent compounds, since the dispersions of the spin fluctuations of high-temperature superconductors are in general analogues to the spin waves of their magnetically ordered parent compounds [2]. For example, the spin fluctuations mainly appear near $(\pi, \pi)$ and $(\pi, 0)$ in cuprate and iron-pnictide superconductors, which can be fitted by a linear spin-wave theory in the Heisenberg model with Néel- and stripe-type magnetic order, respectively [1,2].

Recently, the structurally simplest high-temperature superconductor FeSe has attracted significant interest due to its intriguing superconducting and magnetic properties. The superconducting transition temperature of FeSe is highly tunable; it increases from about 8 K in the bulk form to about 40 K under pressure/electron doping [4-7], and eventually reaches more than 65 K in the single-layer limit [8-15]. Further, the undoped parent phase of FeSe, unlike cuprates and iron pnictides with magnetically ordered parent compounds, undergoes a tetragonal-to-orthorhombic (nematic) transition at $T_s \approx 90$ K but does not exhibit long-range magnetic order down to the lowest measured temperature [16]. The FeSe spin excitation spectrum arises from both the stripe wavevector $(\pi, 0)$ and Néel wavevector $(\pi, \pi)$ and forms a broad continuum at higher energy [17-20], which is also different from simple spin-wave like excitations arising from a single magnetic wavevector in the cuprates $(\pi, \pi)$ and iron pnictides $(\pi, 0)$.

Numerous theoretical models in both the itinerant [21-24] and local moment pictures [25-32] have been used to understand the unusual nematic paramagnetic phase in FeSe. The observed tiny Fermi surface [33,34] and large fluctuating moment [17] imply that the magnetic moments in FeSe are largely localized. Thus, it is reasonable to map the magnetism of FeSe to a spin model in the local moment picture. It has been suggested that the Affleck–Kennedy–Lieb–Tasaki-type nematic quantum paramagnetic phase can be driven by quantum fluctuations of local moments in FeSe [25]. Moreover, density functional theory calculations suggested competing ($\pi$, Q) orders ($-\pi/2 \lesssim Q \lesssim \pi/2$) in FeSe, exhibiting unusual magnetic frustration [26]. Antiferroquadrupolar [27,28] and ferroquadrupolar [29] orders were also proposed to be the ground state of FeSe based on the frustrated bilinear-biquadratic model. In addition, the nematic quantum spin liquid state was also used to describe the inelastic neutron spectra [30,31]. These theoretical proposals are based on models with a specific choice of parameters. Thus, it is crucial to determine the realistic effective magnetic interactions in FeSe. However, owing to the presence of unusual spectra with coexisting Néel- and stripe-type spin fluctuations, and a broad continuum [17], it is difficult to model the spin excitation spectra using simple linear spin-wave theory with a hypothetical magnetic structure. Therefore, a model entirely driven by experimental data without an assumption on the specific magnetic structure is needed to determine the effective exchange interactions in FeSe.

In this paper, we report the neutron scattering measurements and self-consistent Gaussian approximation (SCGA) modeling of the spin excitation spectrum of FeSe in the tetragonal paramagnetic phase. The recently developed SCGA method does not rely on specific magnetic structures, and is successful in determining the magnetic interactions in frustrated magnets in paramagnetic states [35,36]. Our SCGA simulation shows an excellent agreement with the equal-time magnetic structure factor, which reveals highly frustrated nearest-neighbor ($J_1$) and next-nearest-neighbor ($J_2$) exchange couplings in FeSe. These results explain the nematic paramagnetic state in FeSe and provide important insights into the origin of its highly tunable superconductivity.

SCGA is rigorous in the limit of infinite spin dimensions and is a good approximation for the Heisenberg model that is isotropic in spin space [35,37,38]. We first investigate the spin-space anisotropy above the nematic phase transition temperature in FeSe using the spin-polarized neutron scattering technique. The polarized inelastic neutron scattering experiment was conducted on the IN20 thermal neutron triple-axis spectrometer at the Institut Laue-Langevin, France. The FeSe single crystals were co-aligned in the ($H0L$) scattering plane. A Heusler monochromator and analyzer were used to polarize the incident neutron beam and analyze the scattered neutron polarization. The XYZ coordinate system for polarization analysis is defined in the lab frame, where the $x$//$\mathbf{Q}$, and the $z$ axis is in the vertical direction. The incident neutron polarization was aligned parallel ($x$ polarization) or perpendicular ($y$, $z$ polarization) to the momentum transfer, $\mathbf{Q}$. In this configuration, the coherent signal in the spin-flip (SF) channel originates from magnetic scattering, and each component of spin fluctuations perpendicular to $\mathbf{Q}$ ($M_y$, $M_z$) can be obtained from the SF signal with $x$, $y$, $z$ polarization ($SF_x$, $SF_y$, and $SF_z$). A pyrolytic graphite filter was installed in front of the analyzer to eliminate contamination from higher-order neutrons. Fig. 1(c) shows energy-scans at the in-plane wavevector $\mathbf{Q}$ = (1, 0, 0) with different neutron polarized directions at 110 K. No difference between $SF_y$ and $SF_z$ signals was observed below 14 meV [Fig. 1(c)], which is consistent with previous low-energy measurements near $\mathbf{Q}$ = (1, 0, 0) below 8 meV [39]. We also preformed similar polarization analysis at $\mathbf{Q}$ = (1, 0, 1) with out-of-plane component, which again shows that $SF_y$ and $SF_z$ are isotropic [Fig. 1(d)]. The background signal (BG) is derived from $BG = SF_y + SF_z - SF_x$. We obtained the energy dependence of $M_y$ and $M_z$ at $\mathbf{Q}$ = (1, 0, 0) and $\mathbf{Q}$ = (1, 0, 1) [Fig. 1(e)] using the relation $M_y = SF_z - BG$ and $M_z = SF_y - BG$. Two components of spin fluctuations perpendicular to $\mathbf{Q}$ at both (1, 0, 1) and (1, 0, 0) exhibit no difference throughout the measured energy range, demonstrating that excitations in FeSe at 110 K are essentially isotropic in spin space. In addition, the lack of $L$ modulations observed from the neutron scattering experiment [17] indicates the in-plane exchange interactions dominate the magnetic behavior of FeSe. Therefore, the magnetism in FeSe at 110 K can be mapped to the Heisenberg model with $J_1$-$J_2$-$J_3$ in-plane exchange interactions [Fig. 1(a)], and these interaction parameters may be extracted by fitting the spin excitation spectra to the equal-time structure factor calculated using SCGA.

In SCGA, the rigid constraint on the length of individual classical spins $|\mathbf{S}_i|^2 = 1$ is relaxed to the soft one $\langle|\mathbf{S}_i|^2\rangle = 1$. Because different spin components are uncorrelated in the Heisenberg spin system, the partition function can be treated independently for each spin component; thus, the softened spin configurations are weighted by $e^{-\beta \mathcal{H}}$, where

$$\beta \mathcal{H} = \frac{1}{2}\sum_{ij}\left(\lambda \delta_{ij} + \beta \sum_n J_n A_{ij}^{(n)}\right) s_i s_j. \tag{1}$$

$\beta = \frac{1}{k_B T}$, $s_i$ denotes one component of the spin vector $\mathbf{S}_i$, and $A^{(n)}$ is the adjacency matrix between the $n$th nearest neighbors. The Lagrange multiplier, $\lambda$, is determined from the soft spin-length constraint, $\langle s_i^2 \rangle = 1/3$. This constraint fixes the total fluctuating moment, which is consistent with the nearly temperature-independent fluctuating moment in FeSe [17]. For a Bravais lattice, the equal-time spin correlation function in reciprocal space is given as follows:

$$\langle s(\mathbf{q})s(-\mathbf{q})\rangle = \left[\lambda + \beta \sum_n J_n A^{(n)}(\mathbf{q})\right]^{-1}, \tag{2}$$

where $\lambda$ is obtained from the solution of the self-consistent equation

$$\frac{1}{N}\sum_{\mathbf{q}\in BZ}\left[\lambda + \beta \sum_n J_n A^{(n)}(\mathbf{q})\right]^{-1} = \frac{1}{3}, \tag{3}$$

where $N$ is the total number of sites. Then the equal-time structure factor $\mathcal{S}(\mathbf{Q}) = \frac{2}{3}\langle s(\mathbf{Q})s(-\mathbf{Q})\rangle$, can be calculated with the help of Eq. (2).

We used the ARCS time-of-flight chopper spectrometer at the Spallation Neutron Source of Oak Ridge National Laboratory to measure spin correlations in FeSe over a wide momentum range. The energy integrated magnetic intensity, $I(\mathbf{Q})$, was obtained after subtracting the phonon background and incoherent signal. According to the zeroth-moment sum rule [40], the form-factor-corrected $I(\mathbf{Q})$ is proportional to the equal-time magnetic structure factor $\mathcal{S}(\mathbf{Q})$, which can be calculated via SCGA for a given spin Hamiltonian in the paramagnetic state.

We performed a global fit to $I(\mathbf{Q})$ and obtained the optimized parameters $k_B T/J_1 = 0.86 \pm 0.35$, $J_2/J_1 = 0.413 \pm 0.051$ and $J_3/J_1 = 0.069 \pm 0.060$ in the $J_1$-$J_2$-$J_3$ Heisenberg model. Fig. 2 shows the goodness-of-fit $\chi^2$ obtained from $\mathcal{S}(\mathbf{Q})$ fitting at $k_B T/J_1 = 0.86$. The Hessian matrix of $\chi^2$ was used to determine the parameter uncertainties. We note that the fitted magnetic interactions are close to $J_2/J_1 \approx 0.5$ and $J_3/J_1 \approx 0$, which are located at the phase boundary between the Néel and stripe order in the classical mean-field phase diagram of the $J_1$-$J_2$-$J_3$ Heisenberg model [26]. Similar phase boundary or intermediate phase has also been revealed near $J_2/J_1 \approx 0.5$ and $J_3/J_1 \approx 0$ in the spin-1 model by the exact diagonalization (ED) [25,32] and density-matrix renormalization-group (DMRG) [41] calculations.

Fig. 3(a) illustrates the momentum dependence of the structure factor $\mathcal{S}(\mathbf{Q})$ in FeSe, which exhibits a diamond shape and covers a wide region of the Brillouin zone. The $\mathcal{S}(\mathbf{Q})$ calculated from SCGA with the optimized interaction parameters is in good agreement with the data [Fig. 3(b)]. To quantitatively illustrate the structure factor $\mathcal{S}(\mathbf{Q})$, Fig. 4 shows the momentum cuts along the high symmetry directions in FeSe. The SCGA calculations (red lines) again are in excellent agreement with the data. For the optimally fitted parameters, the first moment $K(\mathbf{Q})$ calculated from SCGA is also consistent with the neutron data (see Fig. S1 in Supplemental Material), which further verifies the validity of this method.

We note that the fitted ratio $k_B T/J_1 = 0.86 \pm 0.35$ in SCGA is apparently larger than the realistic value at 110 K. This suggests that quantum fluctuations are not neglectable in FeSe, which easily broaden the spectrum and make SCGA overestimate the temperature. In addition, the presence of itinerant electrons could also induce extra broadening of the spectrum. These effects will influence the absolute value of $k_BT$, but are not expected to affect the ratio of $J_2/J_1$ or $J_3/J_1$ [35,36]. Since $J_3$ is rather weak, the magnetic frustration in FeSe is dominated by the competition between $J_1$ and $J_2$. The $J_2/J_1$ close to 0.5 may naturally account for the observed low-energy Néel and stripe fluctuation at 110 K [17]. The DMRG study [41] revealed an intermediate paramagnetic region (0.525 ≲ $J_2/J_1$ ≲ 0.555) in the phase diagram of the spin-1 $J_1$-$J_2$ model in the zero temperature limit, when the nearest-neighbor exchange interactions along the *a* axis ($J_{1a}$) and *b* axis ($J_{1b}$) are isotropic. The lower boundary of the paramagnetic region will shift down when $J_{1a} \neq J_{1b}$, which is the case for FeSe in the nematic phase at low temperature [41]. Therefore, on cooling to below $T_s$, the magnetic interaction in FeSe settles in the paramagnetic region, where the magnetic frustration and quantum fluctuation prevent the formation of the static magnetic order [25]. The strongly frustrated magnetic interactions near the phase boundaries in FeSe lead to highly tunable magnetism and nematicity under pressure/substrate strain [42,43], which, in turn, could result in highly tunable superconductivity [4,8-15]. Further quantitative measurements of the magnetic interactions in FeSe-based systems with drastically enhanced superconductivity will be particularly interesting.

In summary, we present neutron scattering measurements and SCGA modeling of the spin excitation spectrum of FeSe in the tetragonal paramagnetic phase. The equal-time magnetic structure factor of FeSe exhibits a diamond shape and covers a wide region of the Brillouin zone, which can be quantitatively modeled by SCGA calculation with highly frustrated nearest-neighbor ($J_1$) and next-nearest-neighbor ($J_2$) exchange couplings. This explains the absence of magnetic order in FeSe. The frustrated magnetic exchange interactions could also account for the highly tunable nematic and superconducting properties in FeSe-based superconductors.

This work is supported by the Innovation Program of Shanghai Municipal Education Commission (Grant No. 2017-01-07-00-07-E00018), the National Natural Science Foundation of China (Grant No. 11874119), and the Shanghai Municipal Science and Technology Major Project (Grant No. 2019SHZDZX01).

*zhaoj@fudan.edu.cn

[1] D. J. Scalapino, Rev. Mod. Phys. **84**, 1383 (2012).

[2] P. Dai, Rev. Mod. Phys. **87**, 855 (2015).

[3] Q. Si, R. Yu, and E. Abrahams, Nat. Rev. Mater. **1**, 16017 (2016).

[4] S. Medvedev *et al.*, Nat. Mater. **8**, 630 (2009).

[5] Y. Miyata, K. Nakayama, K. Sugawara, T. Sato, and T. Takahashi, Nat. Mater. **14**, 775 (2015).

[6] X. F. Lu *et al.*, Nat. Mater. **14**, 325 (2015).

[7] E. Dagotto, Rev. Mod. Phys. **85**, 849 (2013).

[8] Q.-Y. Wang *et al.*, Chin. Phys. Lett. **29**, 037402 (2012).

[9] S. He *et al.*, Nat. Mater. **12**, 605 (2013).

[10] S. Tan *et al.*, Nat. Mater. **12**, 634 (2013).

[11] J. J. Lee *et al.*, Nature (London) **515**, 245 (2014).

[12] R. Peng *et al.*, Nat. Commun. **5**, 5044 (2014).

[13] J. F. Ge, Z. L. Liu, C. Liu, C. L. Gao, D. Qian, Q. K. Xue, Y. Liu, and J. F. Jia, Nat. Mater. **14**, 285 (2015).

[14] Y. Xu *et al.*, Nat. Commun. **12**, 2840 (2021).

[15] Y. Song *et al.*, Nat. Commun. **12**, 5926 (2021).

[16] T. M. McQueen, A. J. Williams, P. W. Stephens, J. Tao, Y. Zhu, V. Ksenofontov, F. Casper, C. Felser, and R. J. Cava,

Phys. Rev. Lett. **103**, 057002 (2009).

[17] Q. Wang *et al.*, Nat. Commun. **7**, 12182 (2016).

[18] Q. Wang *et al.*, Nat. Mater. **15**, 159 (2016).

[19] M. C. Rahn, R. A. Ewings, S. J. Sedlmaier, S. J. Clarke, and A. T. Boothroyd, Phys. Rev. B **91**, 180501(R) (2015).

[20] T. Chen *et al.*, Nat. Mater. **18**, 709 (2019).

[21] Y. Yamakawa, S. Onari, and H. Kontani, Phys. Rev. X **6**, 021032 (2016).

[22] A. V. Chubukov, M. Khodas, and R. M. Fernandes, Phys. Rev. X **6**, 041045 (2016).

[23] A. Kreisel, B. M. Andersen, and P. J. Hirschfeld, Phys. Rev. B **98**, 214518 (2018).

[24] X. Zheng, Z. Huang, H. Li, F. Yang, and H. Lin, J. Phys.: Condens. Matter **31**, 055601 (2019).

[25] F. Wang, S. A. Kivelson, and D.-H. Lee, Nat. Phys. **11**, 959 (2015).

[26] J. K. Glasbrenner, I. I. Mazin, H. O. Jeschke, P. J. Hirschfeld, R. M. Fernandes, and R. Valentí, Nat. Phys. **11**, 953 (2015).

[27] R. Yu and Q. Si, Phys. Rev. Lett. **115**, 116401 (2015).

[28] H. H. Lai, W. J. Hu, E. M. Nica, R. Yu, and Q. Si, Phys. Rev. Lett. **118**, 176401 (2017).

[29] Z. Wang, W. J. Hu, and A. H. Nevidomskyy, Phys. Rev. Lett. **116**, 247203 (2016).

[30] J. H. She, M. J. Lawler, and E. A. Kim, Phys. Rev. Lett. **121**, 237002 (2018).

[31] S.-S. Gong, W. Zhu, D. N. Sheng, and K. Yang, Phys. Rev. B **95**, 205132 (2017).

[32] A. Baum *et al.*, Commun. Phys. **2**, 14 (2019).

[33] K. Nakayama, Y. Miyata, G. N. Phan, T. Sato, Y. Tanabe, T. Urata, K. Tanigaki, and T. Takahashi, Phys. Rev. Lett. **113**, 237001 (2014).

[34] M. D. Watson *et al.*, Phys. Rev. B **91**, 155106 (2015).

[35] P. H. Conlon and J. T. Chalker, Phys. Rev. B **81**, 224413 (2010).

[36] K. W. Plumb *et al.*, Nat. Phys. **15**, 54 (2018).

[37] H. E. Stanley, Phys. Rev. **176**, 718 (1968).

[38] D. A. Garanin, Phys. Rev. B **53**, 11593 (1996).

[39] M. Ma, P. Bourges, Y. Sidis, Y. Xu, S. Li, B. Hu, J. Li, F. Wang, and Y. Li, Phys. Rev. X **7**, 021025 (2017).

[40] I. A. Zaliznyak and S.-H. Lee, in *Modern Techniques for Characterizing Magnetic Materials*, edited by Y. Zhu (Springer, Heidelberg, 2005).

[41] H. C. Jiang, F. Krüger, J. E. Moore, D. N. Sheng, J. Zaanen, and Z. Y. Weng, Phys. Rev. B **79**, 174409 (2009).

[42] M. Bendele *et al.*, Phys. Rev. Lett. **104**, 087003 (2010).

[43] J. Pelliciari *et al.*, Nat. Commun. **12**, 3122 (2021).

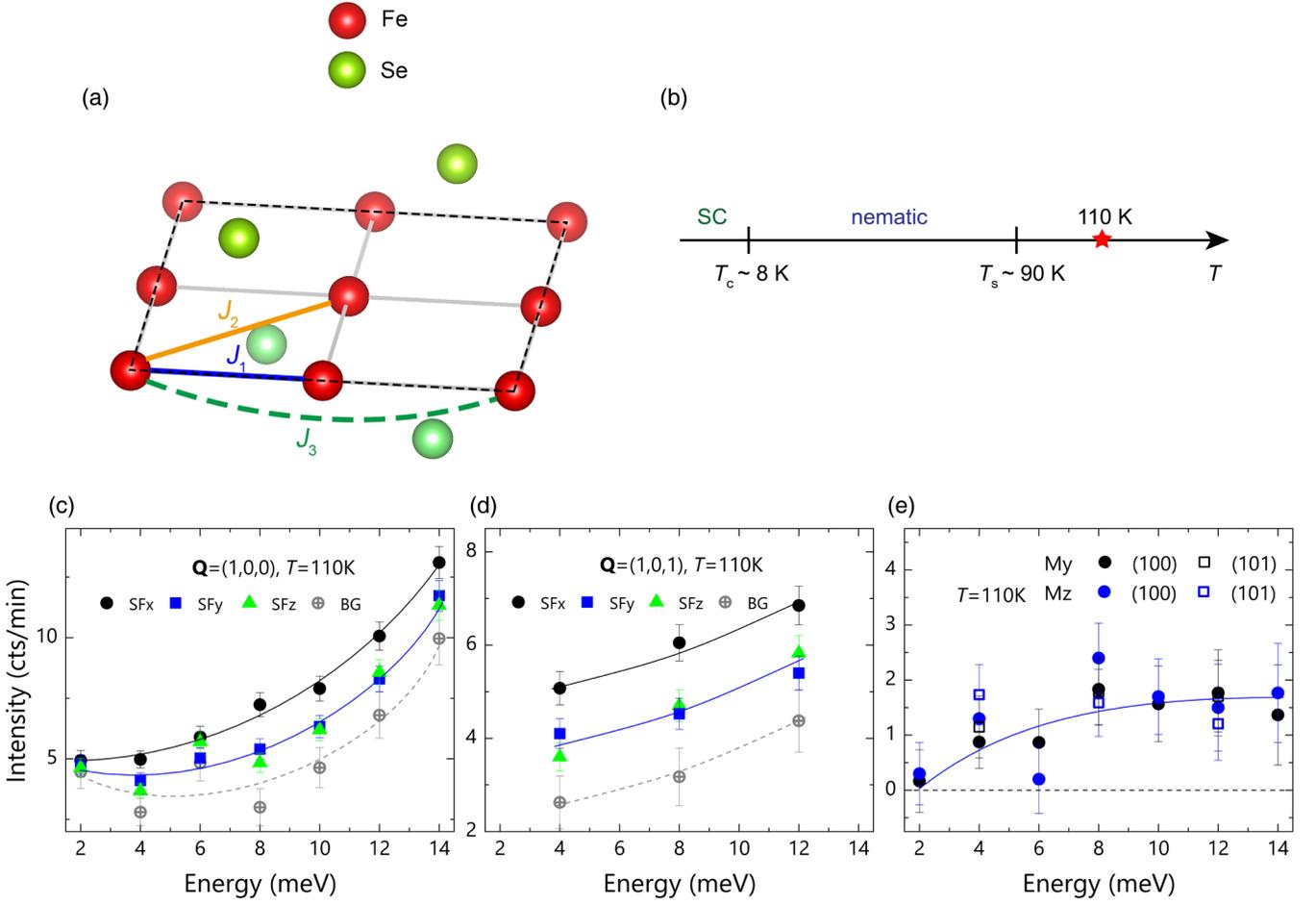

FIG. 1. Crystal structure, phase diagram, and polarized neutron scattering data of FeSe. (a) Crystal structure of FeSe. $J_1$, $J_2$, and $J_3$ denote the nearest-neighbor, next-nearest-neighbor and third-nearest-neighbor in-plane exchange couplings, respectively. The black dashed lines represent the orthorhombic (4-Fe) unit cell projected in *ab*-plane (*a*/*b* axis is along the nearest Fe-Fe bond), which is used throughout our presentation. (b) Phase diagram of FeSe. The red star emphasizes that the magnetic interactions are extracted from the inelastic neutron spectrum measured at 110 K [17]. SC, superconductivity. (c-e) Polarized inelastic neutron scattering data of FeSe at 110 K. The polarization analysis is based on $x$//$\mathbf{Q}$ configuration. Scattering intensities in the spin-flip channel with $x$, $y$, $z$ polarization (SF$_x$, SF$_y$, and SF$_z$) were measured at $\mathbf{Q} = (1, 0, 0)$ (c) and $\mathbf{Q} = (1, 0, 1)$ (d). The background signal (BG) is derived from $BG = SF_y + SF_z − SF_x$. Energy dependence of spin fluctuations $M_y$ and $M_z$ at $\mathbf{Q} = (1, 0, 0)$ and $(1, 0, 1)$ is thus obtained using the relationship, $M_y = SF_z − BG$ and $M_z = SF_y − BG$, with the $Fe^{2+}$ magnetic form factor corrected (e). Error bars indicate one standard deviation. The solid and dashed lines are a guide to the eye.

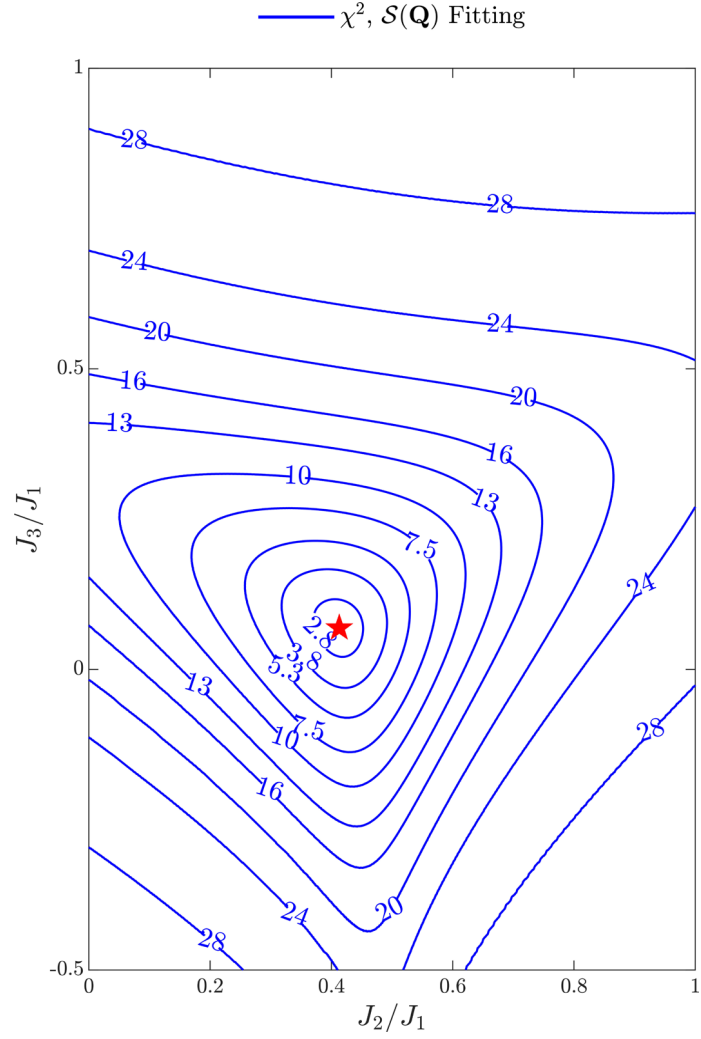

FIG. 2. Determination of magnetic interactions in FeSe. Contour plot of the goodness of fit $\chi^2$ between calculations and neutron scattering data. The equal-time magnetic structure factor, $\mathcal{S}(\mathbf{Q})$, is calculated through the self-consistent Gaussian approximation for the $J_1$-$J_2$-$J_3$ Heisenberg model at $k_B T/J_1 = 0.86$. Inelastic neutron scattering data of FeSe were collected at 110 K [17]. The red star denotes the best fit interaction parameters.

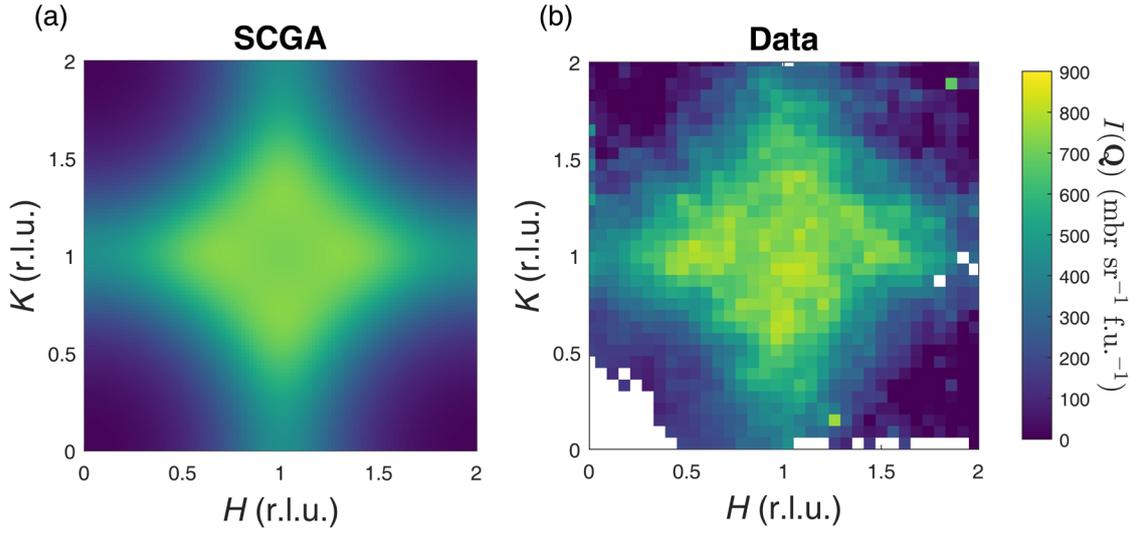

FIG. 3. Momentum dependence of calculated and measured equal-time structure factor in FeSe. (a) The equal-time structure factor calculated using SCGA for the optimized parameters $k_B T/J_1 = 0.86$, $J_2/J_1 = 0.413$ and $J_3/J_1 = 0.069$. (b) The energy-integrated intensity, $I(\mathbf{Q}) = \int_0^{E'}(1 + e^{-E/k_B T})I(\mathbf{Q},E)dE$, obtained from the measured magnetic intensity, $I(\mathbf{Q},E)$, of FeSe at $T = 110$ K, with the $Fe^{2+}$ magnetic form factor corrected. $E' = 220$ meV is the spin excitation energy's upper limit. The wavevector $\mathbf{Q}$ at $(q_x, q_y, q_z)$ is defined as $(H, K, L) = (q_x a/2\pi, q_y b/2\pi, q_z c/2\pi)$ in reciprocal lattice units (r.l.u.) of the orthorhombic unit cell, where $a = b = 2d_{Fe-Fe}$ is twice the nearest Fe-Fe distance in the tetragonal phase.

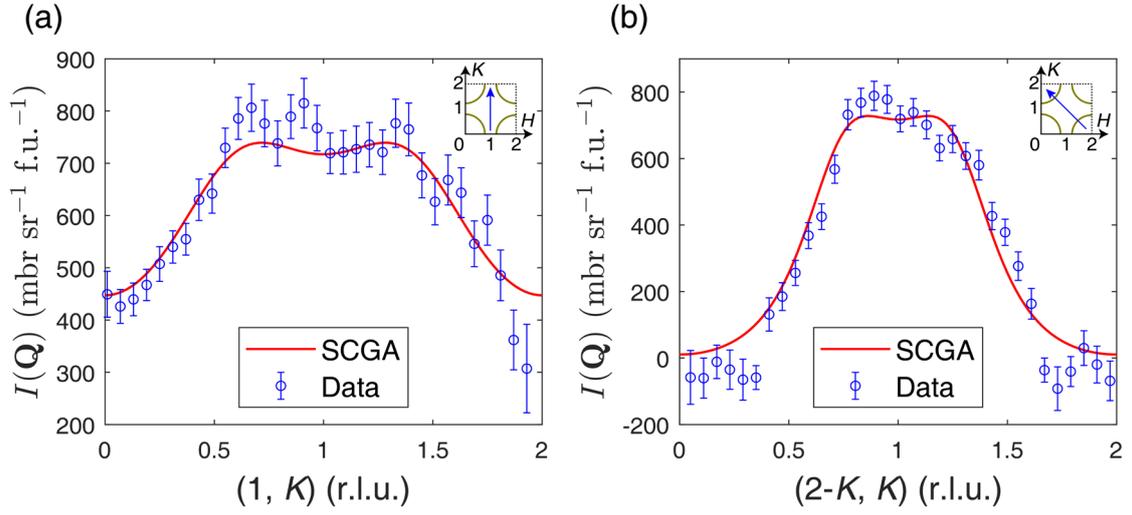

FIG. 4. Calculated and measured equal-time structure factor along the high symmetry directions in FeSe. (a, b) Measured equal-time structure factor along the high symmetry directions and comparison with SCGA calculations for the optimized parameters $k_B T/J_1 = 0.86$, $J_2/J_1 = 0.413$ and $J_3/J_1 = 0.069$. The scan directions are marked by the arrows in the insets. Error bars indicate one standard deviation.